\documentclass[iop]{emulateapj}

\usepackage{graphicx}
\usepackage{amsmath}
\usepackage{subfigure}
\usepackage{float}
\usepackage{natbib} 
\usepackage{hyperref}

\begin{document}

\title{EvryFlare. IV. Detection of periodicity in flare occurrence from cool stars with TESS}

\author{Ward S. Howard\altaffilmark{1}, Nicholas M. Law\altaffilmark{1}}

\altaffiltext{1}{Department of Physics and Astronomy, University of North Carolina at Chapel Hill, Chapel Hill, NC 27599-3255, USA}

\email[$\star$~E-mail:~]{wshoward@unc.edu}

\begin{abstract}
Phased flaring, or the periodic occurrence of stellar flares, may probe electromagnetic star-planet interaction (SPI), binary interaction, or magnetic conditions in spots. For the first time, we explore flare periodograms for a large sample of flare stars to identify periodicity due to magnetic interactions with orbiting companions, magnetic reservoirs, or rotational phase. Previous large surveys have explored periodicity at the stellar rotation period, but we do not assume periods must correspond with rotation in this work. Two min TESS light curves of 284 cool stars are searched for periods from 1-10 d using two newly-developed periodograms. Because flares are discrete events in noisy and incomplete data, typical periodograms are not well-suited to detect phased flaring. We construct and test a new Bayesian likelihood periodogram and a modified Lomb-Scargle periodogram. We find 6 candidates with a false-alarm probability below 1\%. Three targets are $\geq$3$\sigma$ detections of flare periodicity; the others are plausible candidates which cannot be individually confirmed. Periods range from 1.35 to 6.7 d and some, but not all, correlate with the stellar rotation period or its 1/2 alias. Periodicity from 2 targets may persist from TESS Cycle 1 into Cycle 3. The periodicity does not appear to persist for the others. Long-term changes in periodicity may result from the spot evolution observed from each candidate, which suggests magnetic conditions play an important role in sustaining periodicity.
\end{abstract}

\maketitle

\section{Introduction}
M-dwarf flares are generally thought to be stochastic events \citep{hawley2014}. There are exceptions, however. Over the years, various authors have searched for periodicity in the timing between flare events with various degrees of success (e.g. \citealt{Kunkel1971, Pettersen1983, Dal2011a, Doyle2018, Roettenbacher2018}). If present, flare periodicity would serve as a unique probe of stellar rotation, star-planet interaction, or the surface magnetic properties of starspot groups. The possible explanations of apparent flare periodicity generally fall into four categories: rotating starspots, magnetic interaction with a companion star or planet, magnetic ``reservoir hypotheses," and random chance.

The rotation of spotted stars can induce regular variability in light curves. Stellar activity present in the photosphere and chromosphere of M-dwarf stars are frequently coupled to one another (e.g. \citealt{Korhonen2010, Vida2017} and references therein). Flares associated with these spots may therefore preferentially occur at a certain rotational phase (such as when the dominant spot group is facing us). Preferentially-phased flares have been reported for individual stars (e.g. \citealt{Mavridis1995, Dal2011a, Dal2011b, Pi2019, Ioannidis2020}). Other observations show little to no preference for flaring at particular rotational phases for individual stars (e.g. \citealt{Lukatskaia1976, Pettersen1983, Hunt_Walker2012, Roettenbacher2013, Vida2017}). Non-detections may be due to unfavorable spot geometries such as polar spots or spots across all longitudes \citep{Roettenbacher2018, Feinstein2020, Ilin2021}. Larger surveys have also reported mixed results. For example, \citet{Roettenbacher2018} do find a correlation of flares with spots in a sample of 119 late F to early M stars observed during the primary \textit{Kepler} mission \citep{Borucki2010}. However, preferential flaring with rotational phase was not supported in a study of 34 flaring M-dwarfs \citep{Doyle2018} observed during the K2 mission \citep{Howell2014} or 119 flaring M-dwarfs observed in TESS Sectors 1 through 3 \citep{Doyle2019}. \citet{Feinstein2020} do not find evidence for a correlation of spots and flaring in a sample of 1500 young rotators in 2 min TESS data, hypothesizing spots that cover the entire photosphere.

Binary companions or exoplanets in close orbits around M-dwarfs may induce flares and otherwise alter the stellar atmosphere via electromagnetic interactions, e.g. \citealt{Ip2004, Preusse2007, Gao2008, Strugarek2015, Lanza2018}. \citet{Doyle1990} observed a clear 48 min periodicity in 4 large flares from the eclipsing binary YY Gem with a false positive probability of $<$0.5\%. \citet{Gao2008} were able to reproduce this periodicity via a star-star interaction. Star-star interactions have also been proposed as an explanation for phased flaring from V711 Tau \citep{Garcia_Alvarez2003} and UX Ari \citep{Cao2017}.

Planets may also induce regular chromospheric changes by depositing magnetic energy in the stellar atmosphere. Electromagnetic star-planet interactions (SPIs) may occur when the convection velocity of the stellar wind is sub-Alfv\'enic, coupling the star and planet and enabling efficient transfer of energy \citep{Goertz1980,Neubauer1980,Fischer_Saur2019}. SPIs may also occur when the star and planet couple via magnetohydrostatic force-free magnetic fields \citep{Lanza2009,Lanza2012, Lanza2013, Lanza2018}. Most electromagnetic SPIs are observed as periodic changes to emission lines or photometry and are due to close-in giant planets. Examples include HD 179949 and $\nu$ And \citep{Shkolnik2003,Shkolnik2005,Shkolnik2008},  $\tau$ Boo A \citep{Walker2008}, WASP-43 \citep{Staab2017}, HD 189733 \citep{Cauley2018}, and Gliese 1151 \citep{Vedantham2020, Mahadevan2021} (Recently, RV data from CARMENES has called the existence of the Gliese 1151 planet into doubt \citep{Perger2021}). Flare periodicity due to SPIs rather than stellar companions has a firm theoretical basis in \citet{Lanza2018} but remains unconfirmed with observations. A tentative flaring SPI has been suggested in the TRAPPIST-1 system from interactions with TRAPPIST-1 b and c \citep{Fischer_Saur2019}.

As an alternate source of periodicity, the ``flare reservoir hypothesis" describes the release of stored magnetic energy in an active region. After a flare occurs, some amount of time may be needed to increase the stored energy before another flare of similar energy can be triggered. If the amount of time needed to reach a critical energy remains consistent, flare periodicity would result \citep{Pettersen1983, Doyle1990, Garcia_Alvarez2003}. The reservoir hypothesis is qualitative and may be consistent with a range of physical models \citep{Pettersen1983}. The hypothesis shares similarities with the phenomenon of quasi-periodic pulsations (QPPs) but might operate at longer timescales \citep{Pugh2016}. \citet{Pettersen1983} did not find any evidence for a flare reservoir in observations of 17 flares from V780 Tau. \citet{Garcia_Alvarez2003} found a 2.8 d periodicity in 7 large X-ray flares from V711 Tau with a false positive probability of $\sim$0.5\%. Flares occur twice per stellar rotation, but the preciseness of the event timing suggests the periodicity is not due entirely to rotating starspots and could involve a reservoir \citep{Garcia_Alvarez2003}.

A final explanation of apparent flare periodicity is random chance. When a small number of flares from a star show apparent periodicity, the chance that the signal is due to a random Poisson process must be rejected \citep{Doyle1990, Garcia_Alvarez2003}. A periodicity search across a sample of hundreds to thousands of flare stars such as those observed by Kepler (e.g. \citealt{davenport2016, Doorsselaere2017, Yang2017}) or TESS (e.g. \citealt{Doyle2019, Gunther2020, Feinstein2020}) may also result in false-positive signals. False-positives arise from the look-elsewhere effect or problem of multiple comparisons \citep{Miller1981,Shaffer1995,Bayer2020}. For example, if 330 flare stars are searched for periodicity, at least one 3$\sigma$ detection will result by chance alone.

\subsection{Surveying stars for flare periodicity}\label{flare_survey_tools}
In its initial and extended missions, TESS \citep{Ricker_TESS} is providing an unprecedented window into M-dwarf flare activity \citep{Gunther2020}. At least 15,000 of the brightest late K and M-dwarfs have been observed at 2 min cadence for at least 1 sector \citep{Stassun2019}, enabling comprehensive searches for flare periodicity from individual stars across a wide sample \citep{Feinstein2020}. Previous large-scale periodicity surveys have only explored periods at the stellar rotation period, e.g. \citealt{Roettenbacher2018, Doyle2019, Feinstein2020}.  These surveys primarily explore periodicity on a statistical rather than case-by-case basis, although \citet{Doyle2019} also inspected 45 flare stars for rotational modulation of flares. We develop and test new flare periodogram tools on a sub-sample of 284 very active and well-studied flare stars from \citet{Howard_2019} as a step toward a broader survey. We follow the insight of transit surveys designed to detect periodicity that may only be present in a small fraction of stars. to avoid diluting strong periodic signals from individual stars with non-periodicity from other stars, we search for flare periodicity star by star.

\subsection{Tools for detecting periodicity from discrete events in noisy data}\label{previous_period_tools}
Flares are discrete and stochastic events, motivating periodograms for incomplete, noisy data. Discrete Fourier Transforms, auto-correlation approaches, histograms of arrival time separations, and string comparison methods are not designed for incomplete or pseudo-periodic event times \citep{Li2015, Ghosh2017}. \citet{Li2015} and \citet{Zhang2018} create periodograms employing probabilistic likelihoods for various test periods. Whenever the test period approaches a real periodicity, the phase-folded points line up in phase and create high likelihoods. \citet{Ghosh2017} searched for periodicity in incomplete data with Bayesian particle bootstrap filters. They model periodicity with incomplete data and where the preferred phase is not fixed in time. We present a new Bayesian likelihood periodogram for discrete events in noisy and incomplete data. We also present a modified Lomb-Scargle periodogram for comparison.  Our Bayesian periodogram is similar to the \citet{Li2015} approach but is fully Bayesian, making it an efficient and robust way to characterize flare signals that can easily incorporate priors. Our approach is simpler than the bootstrap filter since we do not need to perform online prediction, but rather to search many stars.

\begin{figure*}
	\centering
	{
		\includegraphics[trim= 1 20 1 10,clip, width=6.9in]{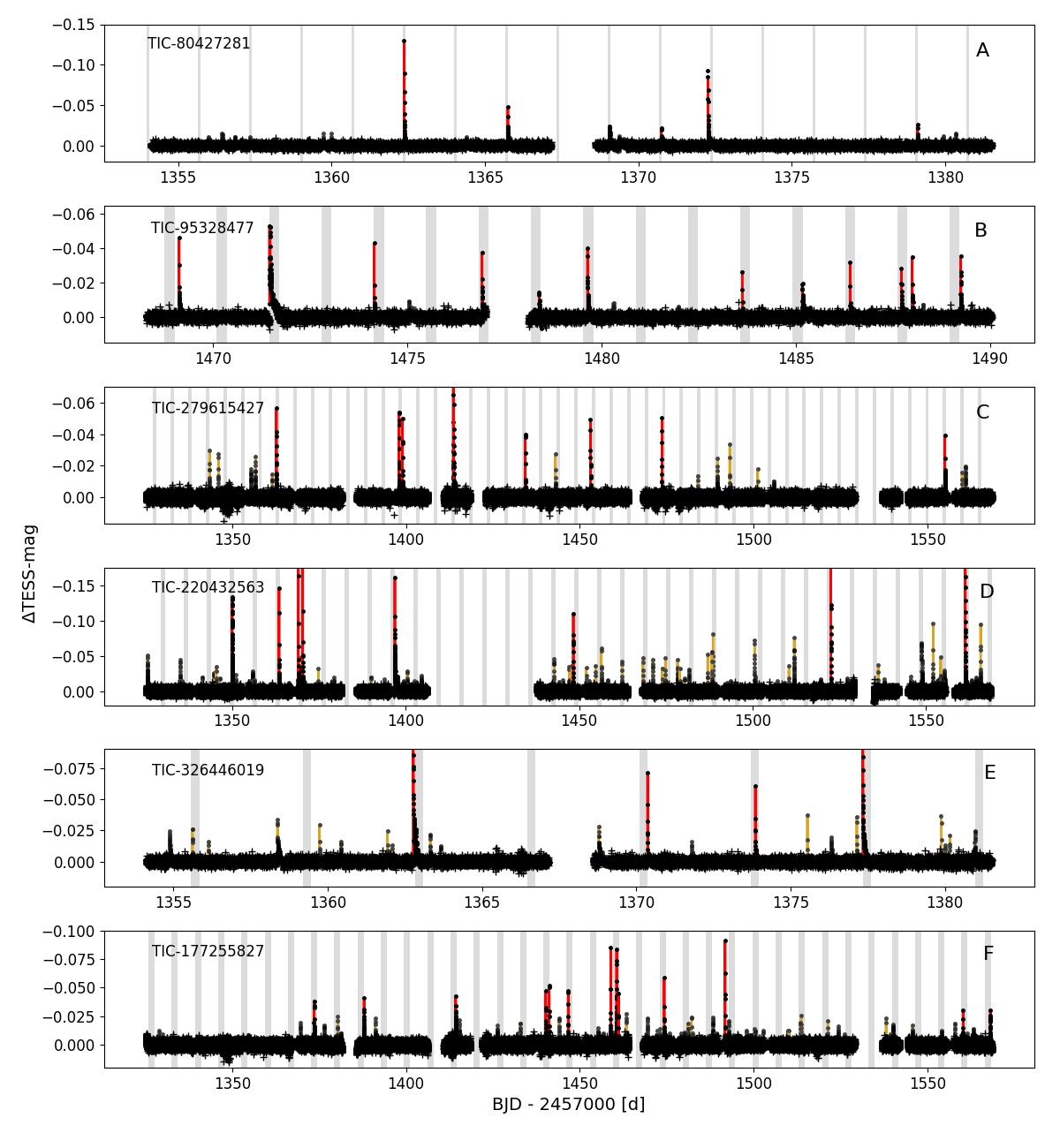}
	}
	\caption[TESS light curves of the 6 candidate phased-flares targets]{TESS Cycle 1 light curves of the 6 candidate phased-flares targets with FAP$<$1\%. Periodicity from these flare stars was detected with our newly-developed flare periodograms. Flares large enough to comprise the phased-flares signal (i.e. with amplitudes above the threshold amplitude $A_\mathrm{thresh}$) are highlighted in red. Smaller flares are highlighted in light orange. Predicted times when high-amplitude flares will occur are shaded in grey. Out-of-flare variability in the light curve is de-trended with a Savitsky-Golay filter. Some phased-flare targets are stronger detections than others, with A and B as the clearest.}
	\label{fig:candidate_lightcurves_fig}
\end{figure*}

In this work, we construct phased-flares periodograms and explore their performance detecting simulated periodicity (\S \ref{methods_periodogram}). Using the periodograms, we perform a large periodicity survey in TESS light curves (\S \ref{survey_results}) and consider the physical mechanisms behind our phased flares signals (\S \ref{stuff1}).

\section{TESS flare observations}\label{tess_observations}
TESS is continuously monitoring each part of the sky for 28 d at a time in the red with four 10.5 cm optical telescopes at 21$\arcsec$ pixel$^{-1}$. During the primary mission, TESS Cycle 1 observed the southern hemisphere split into 13 sectors. TESS Cycle 3 is now re-observing this field as part of the extended mission. We downloaded 2 min cadence simple aperture photometry (SAP) light curves from TESS Sectors 1-9 of 284 late K and M-dwarf flare stars from \citet{Howard_2019}. Following \citet{Vida_2019}, SAP rather than pre-conditioned light curves were selected to avoid altering or removing flares or other astrophysical variability.

We identify flares in the light curves of all targets with the \citet{Howard_2020b} automated pipeline and then again independently in a visual search for 11 targets with apparent periodicity to ensure no flares are missed. In both, flares are discovered as brightening events above the photometric noise following the rapid-rise, slow-decay \citep{davenport2014} flare profile. Flare identification does not require \citep{davenport2014} profiles, although most flares do exhibit morphologies well-described by them (or super-positions of them). More symmetrical events are also included if they are clearly distinguishable from the local photometric noise. Visual identification is performed to ensure the start and stop times of each flare are correct and that no artifact of automated flare detection is the cause of flare periodicity. Out-of-flare variability is removed using a Savitzky–Golay (SG; \citealt{Savitzky1964}) filter, carefully inspecting each flare to ensure out-of-flare variability is correctly subtracted. Carefully removing rotational variability is essential to ensuring flux is not added to or removed from the flares in a periodic way. A SG filter is a procedure using local polynomial regression, or LOESS regression. SG parameters were chosen to give good by-eye agreement to the light curve without over or under fitting. We tune two parameters in the \texttt{scipy} SG implementation: (1) the window length, or the size of the filter window, and (2) the polynomial order of the fit. If the stellar rotation period $P_\mathrm{rot} <$ 1 d, then the window length and polynomial order are set to 101 and 2, respectively. If 1 $\leq P_\mathrm{rot} <$ 2 d, then they are set to 151 and 2. If 2 $\leq P_\mathrm{rot} <$ 4 d, then they are set to 201 and 2. Otherwise, the parameters are set to 401 and 1. All SG values assume a light curve in units of days.

Finally, flare amplitudes are measured in the de-trended light curves using fractional fluxes instead of magnitudes to observe linear relationships between flare peaks. Fractional flux is computed as $\Delta$F/F=$\frac{\mid F-F_0 \mid}{F_0}$ where $F_0$ is the out-of-flare flux and is determined from the local median. Stellar distances and the $T$ magnitude of the star are primarily obtained from the TESS Input Catalog version 8 (TIC; \citealt{Bailer_Jones2018,Stassun2019}). Stellar rotation periods are identified by strong rotational modulation in the TESS light curves in both Cycle 1 and Cycle 3. LS periodograms are computed for the out-of-flare variability and the light curve is folded in phase to confirm the period.

\begin{figure*}
	\centering
	{
		\includegraphics[trim= 1 30 1 15,clip, width=6.9in]{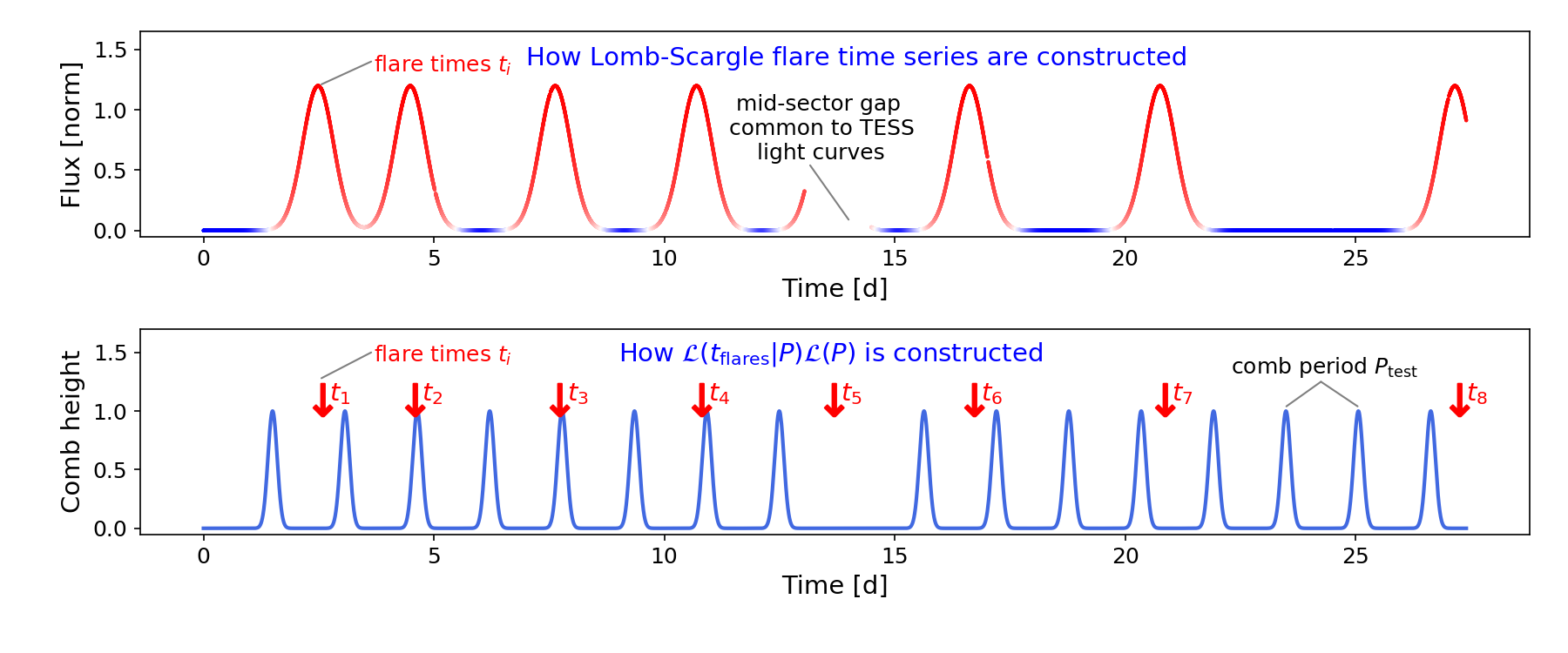}
	}
	\caption[Flare periodogram methods]{Top panel: A TESS light curve modified for input into the wide-flares LS periodogram. Y-values in the light curve are replaced with a small constant value where there are no flares and a Gaussian where there is a flare. The Gaussian is made much wider than the original flare in order to be picked up by the LS basis functions. Bottom panel: The flare probability comb with period $P$ used to compute Bayesian likelihood periodograms $\mathcal{L}$($P | t_i$) = $\mathcal{L}$($t_i | P$)$\mathcal{L}$($P$) for a set of discrete flare times $t_i$. The comb determines $\mathcal{L}$($t_i | P$) as phased-flares with period $P$ occur near the peaks of the comb teeth more often than flares phased to a different period (or no period). The comb is swept through the flare times to account for phase and the maximum likelihood is recorded for each period in the periodogram.}
	\label{fig:methods_explainer}
\end{figure*}

\section{Flare periodicity search methods}\label{methods_periodogram}
We outline the identification of flare periodicity in TESS light curves and the development of two new phased-flares periodograms. Because the Bayesian periodogram is a fundamentally-new approach to flare periodicity, we construct a modified LS periodogram as a way to assess the dependence of signal power on the type of periodogram. While the simplest approach often used to detect periodicity is to compute the distribution of inter-arrival times, we use a more sophisticated approach to account for missing events due to the stochastic nature of stellar flare emission. Flares shouldn't be expected at every predicted occurrence time as the magnetic conditions of starspots may not be that consistent. Stellar rotation will also hide some flares from our line of sight. In the following subsections, we discuss the steps required to measure and assess each periodogram:
\begin{enumerate}
    \item \S \ref{bayesian_PFB}: Generate the Bayesian likelihood periodogram.
    \item \S \ref{LS_methods}: Generate the Wide-flares LS periodogram.
    \item \S \ref{methods_compare}: Assessing the performance of the periodograms as a function of the number of flares.
    \item \S \ref{ID_signal_methods}: Use the periodograms and a visual inspection of each of the 284 light curves to identify candidate phased flare signals for further study. 
    \item \S \ref{ampl_thresh}: Find the threshold amplitude separating large, phased flares from smaller, non-phased flares. This threshold is necessary to select the finalized sample of events from each discovery on which periodograms and false-alarm probabilities will be run.
    \item \S \ref{fap_methods}: Identify strong periodogram signals for all flares above the threshold amplitude for each star using FAP calculations.
\end{enumerate}

\begin{figure*}
	\centering
	{
		\includegraphics[trim= 1 1 1 1,clip, width=6.9in]{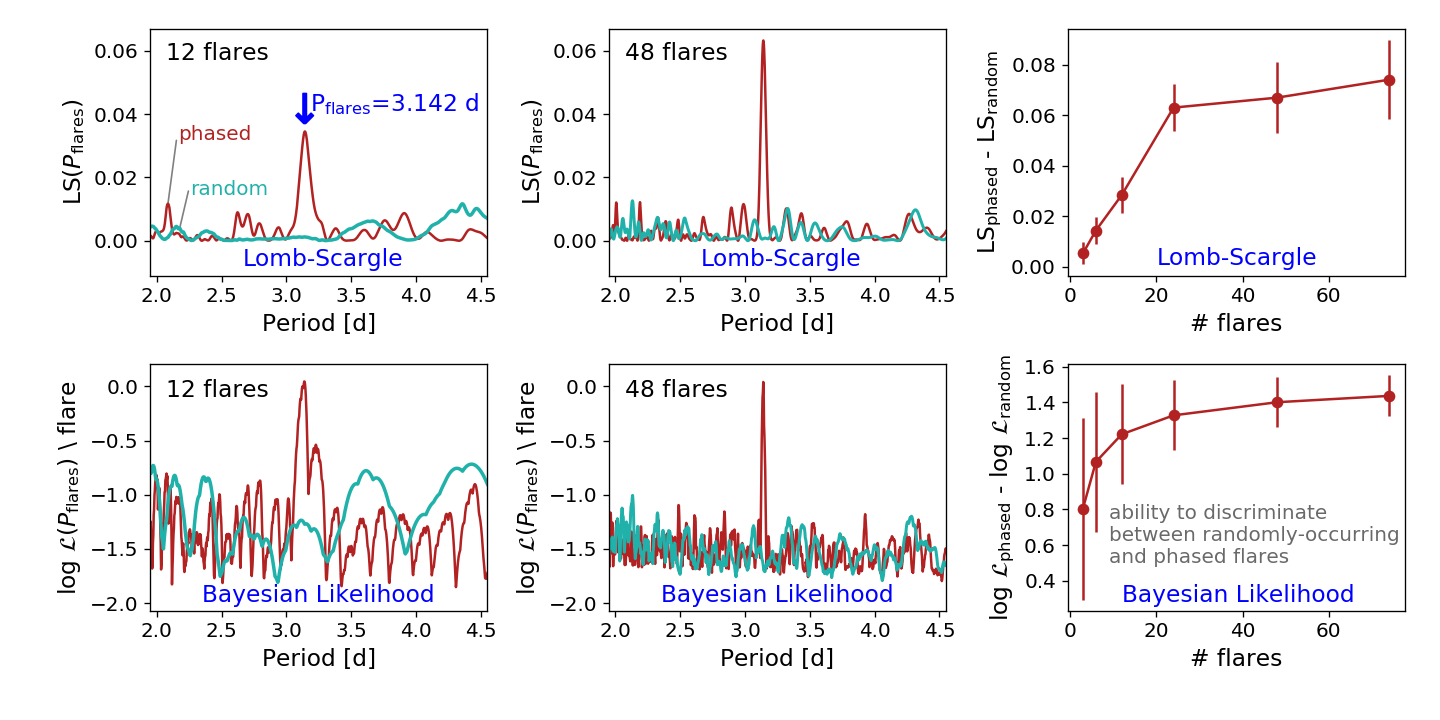}
	}
	\caption[Flare periodogram power vs. number of flares]{Simulated detection of a 3.142 d phased-flares periodicity with the wide-flares LS periodogram (top panels) and Bayesian likelihood periodogram (bottom panels). Periodograms of randomly-timed flares are also shown. The leftmost panels show phased and random flare detections from just 12 events, while the middle panels show phased and random flare periodograms from 48 events. As the number of flares increases, the signal to noise increases and the ability to discriminate phased flares from random flaring also increases. The rightmost panels extend this observation with MC trials on sets of 3, 6, 12, 24, 48, and 74 flares. The Y-axes of the rightmost panels  LS$_\mathrm{phased}$-LS$_\mathrm{random}$ and log $\mathcal{L}_\mathrm{phased}$-log $\mathcal{L}_\mathrm{random}$ gives the typical increase in power between real detections versus non-detections. These differences in power allow us to explore the ability of the periodograms to distinguish between phased and random flaring scenarios as a function of the number of flares. The most improvement is seen from 3-24 flares.}
	\label{fig:snr_demo}
\end{figure*}

\subsection{Phased-flares Bayesian Likelihood periodogram}\label{bayesian_PFB}
We employ Bayes' theorem to to compute the likelihood that a set of discrete stellar flare times demonstrates periodicity at a given period. Computing $\mathcal{L}$($P\vert F$), the likelihood of periodicity at period $P$ for a set of discrete flare times $F$, is more difficult than computing the likelihood that a set of flare times $F$ occur at a period $P$, $\mathcal{L}$($F\vert P$). This is because it is a straightforward matter to compute whether flares occur at high-probability intervals on a periodic probability distribution at a given period $P$. We therefore compute $\mathcal{L}$($P\vert F$) as
\begin{equation}
    \mathcal{L}(P \vert F)=\mathcal{L}(F \vert P) \times \mathcal{L}(P)= \prod_i \mathcal{L}(F_i \vert P)\times \mathcal{L}(P)
    \label{eq:notsu_spot_flare_scaling}
\end{equation}
where $F_i$ is the time at which each discrete flare event occurs and $\mathcal{L}$($P$) is the prior likelihood of each period (i.e. a period of at a known systematic would be unlikely). In this work, we set $\mathcal{L}$($P$) to one for simplicity. In the larger survey, we could use a prior based on the TESS window function, excluding periods at 14 and 28 d, for example. We note setting the prior to one makes our solution equivalent to non-Bayesian maximum likelihood estimation. However, this choice of prior is optional in our periodogram setup.

Flare likelihoods $\mathcal{L}(F_i \vert P)$ are evaluated in $\log \mathcal{L}(t \vert P)$ as the composition of the log of a Gaussian function $G$ centered at zero ($\mu_G$=0; the mean position of the Gaussian) with a comb of delta functions $\sum_{k=-\infty}^{\infty} \delta (t-kP)$.
\begin{equation}
    \log \mathcal{L}(t \vert P) = \log G(t, \mu_G, \sigma_G) * \sum_{k=-\infty}^{\infty} \delta (t-kP) \bigg\rvert_{F_i}
    \label{eq:thenthewingedhussarsarrived}
\end{equation}
Each delta function is spaced $P$ days after the previous delta function in the comb. The delta function builds in the periodicity, while the Gaussian allows for the stochastic nature of flare events emitted before or after the exact center of the probability peak of the comb tooth. The resulting likelihood function consists of a comb of evenly-spaced Gaussian teeth with a separation of $P$ days as shown in the bottom panel of Figure \ref{fig:methods_explainer}. Each Gaussian tooth has a spread of $\sigma_G$=0.1 d about the predicted time. $\sigma_G$=0.1 d was selected as a good compromise between tooth width and minimum test period in the periodogram (see the following discussion on the relationship between $\sigma_G$ and minimum test period). Since reconnection is stochastic, flare events triggered by a periodic mechanism will not occur at exactly periodic intervals, but may be clustered about the expected phase within some interval of time. We account for this pseudo-periodicity by using Gaussian-shaped probability envelopes for the teeth (which makes it most probable the flares occur at exactly the same phase at each period, but admits a range of possible occurrence times). The choice of a Gaussian ensures the periodogram will detect flare periodicity even when the events are not strictly periodic, but vary before or after the expected time by $\sim$0.1 d. The chosen value of $\sigma_G$ has two implications for the final Bayesian periodogram. First, the periodogram will break down at periods $P$ for which $\sigma_G$/$P\geq \sim$1. Second, the amount of random scatter in the flare times around the exact period $P$ that will generate high probabilities is set by $\sigma_G$.

Once a likelihood function $\log \mathcal{L}(P \vert F)$ has been defined for a given period $P$, a periodogram is created across all periods. For each period $P$, the comb is swept through the flare times to account for any mismatch in phase between periodic flares and the placement of the teeth. The highest likelihood is recorded for that test period. The normalization constant of the likelihood scales as $\sim P^{2}$ due to the decrease in the area under the comb at longer periods, similar to the effect seen by \citet{Li2015}. At short periods ($<$2 d), the periodogram power noise floor begins to increase due to the closeness of the comb spacing. This slight excess in power compared to longer periods is well-described by an exponential function and is removed in plots to guide the eye.

Our method shares some superficial similarities to that of \citet{Nava2016} and references therein, which is used to detect periodicity in earthquake events. For example, both methods search for periodicity in a discrete set of event times using a Dirac comb. Both use Gaussians in event prediction to account for imperfect periodicity. Significant difference also exist. For example, \citet{Nava2016} uses a Fourier transform to find periodicity before applying the Dirac comb, while we use the comb and Bayes' Theorem to directly identify candidate periods. While we use Bayes' Theorem in the initial periodicity detection, \citet{Nava2016} use Bayes' Theorem in forecasting events. They also use Bayes' Theorem to update the probability that a sequence of events shows evidence of periodicity after additional events are detected. Finally, while we use a uniform prior, \citet{Nava2016} explore several subjective priors.

\begin{figure*}
	\centering
	{
		\includegraphics[trim= 1 30 1 10,clip, width=6.9in]{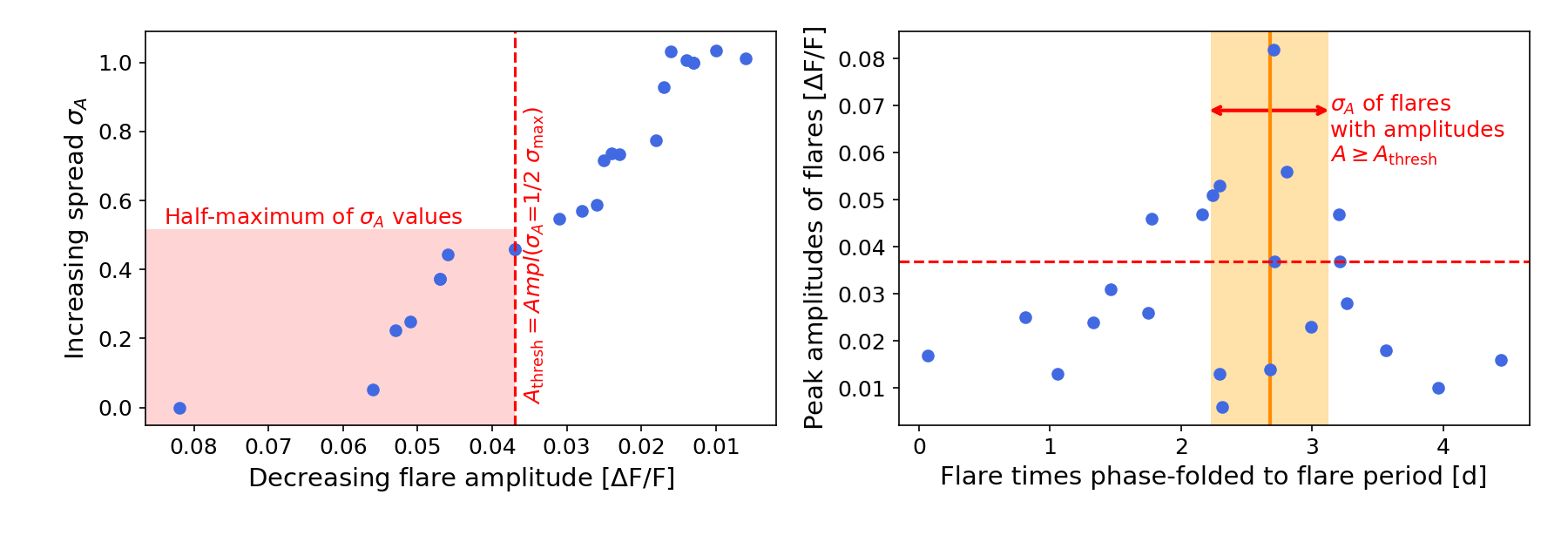}
	}
	\caption[Flare amplitude threshold determination]{The method by which flares from each light curve are sorted into the high-amplitude phased-flares and into the low-amplitude random flares. Because low-amplitude flares often appear uncorrelated with each phased-flares signal, we define a threshold amplitude $A_\mathrm{thresh}$ to isolate the larger flares for periodogram and statistical analysis. In the right panel, we plot flare amplitudes versus the phase-folded flare times. For each flare in the right panel, we measure the spread in phase $\sigma_A$ for all flares with amplitudes $A_\mathrm{flares} \geq A_\mathrm{thresh}$. As the flare amplitude decreases, $\sigma_A$ increases. This creates the $A-\sigma_A$ distribution shown in the left panel. The amplitude $A_\mathrm{thresh}$ is selected to correspond to the half-maximum of $\sigma_A$. For all $N_\mathrm{above}$ flares above $A_\mathrm{thresh}$, the standard deviation of their phase-folded times is computed. Phased-flares FAP are computed for each candidate signal by randomly-shuffling flare times and amplitudes to see how often $N_\mathrm{above}$ random flares can produce a smaller standard deviation than the actual flares.}
	\label{fig:sigma_amplFWHM}
\end{figure*}

\subsection{Wide-flare Lomb-Scargle periodogram}\label{LS_methods}
Lomb-Scargle (LS) periodograms are a standard method of detecting periodicity in a light curve \citep{Scargle1982, VanderPlas2018}. Because they correlate sinusoidal functions with the light curve, they are not automatically a good matched filter for very short-term impulsive events like flares. To make the LS basis functions applicable to periodic flare signals we artificially widen the flare signals by replacing the light curve with a synthetic light curve consisting of a near-zero constant value $C$ where no flares are observed, and wide Gaussian-shaped bumps $G$ where flares $F_i$ do occur. This synthetic light curve $y(t)$ is constructed as the convolution
\begin{equation}
    y(t) = G(t, \mu_G, \sigma_G) * \sum_{i=1}^{N} \delta (t-F_i) + C
    \label{eq:LSinput}
\end{equation}
with $N$ being the number of flares and the Gaussian mean and spread as before. If exactly zero is used for $C$, the window function is not preserved as out-of-flare times no longer contribute to the convolution. The wide-flares process is illustrated in the top panel of Figure \ref{fig:methods_explainer} where Gaussian flux increases at the flare times are highlighted in red and the non-flare times are displayed in blue. The TESS observing window is preserved since only the flux values in the light curve are synthetic, not the epoch times. For example, the mid-sector TESS observing gap can be seen in Figure \ref{fig:methods_explainer}. Once $y(t)$ has been constructed, the normal LS periodogram can be computed as LS($y(t)$). Since LS periodograms are well-understood, they provide a good check on the performance of the new Bayesian likelihood periodogram.

\subsection{Performance of each method}\label{methods_compare}
For a TESS sector of 28 days and a phased-flares period of 1-5 days, between 6-30 periodic flares will occur. How well do the Wide-flares LS and Bayesian-likelihood periodograms distinguish a phased-flares signal from random flaring for so few flares? We first verify both periodograms recover injected flares at periods from 1 to 10 d, with stronger signals for higher numbers of flares as illustrated in the left and middle panels of Figure \ref{fig:snr_demo}. By necessity, both types of periodograms generate power at the primary period and also a signal at the 1/2 alias. More flares are required at 10 d periods than 2 d periods to reach the same peak strength. Injected flares are initially simulated at the precise period, and then scatter is added, drawn from a random normal distribution with a 1$\sigma$ value of 0.05 d.

To determine the typical periodogram power obtained with flares in periodic versus random configurations, 10,000 MC trials are performed on each sample of $N$ random and $N$ phased-flares. We separately test samples of $N$=3, 6, 12, 24, 48, and 74 random flares; we test the same numbers of phased-flares. As expected, larger numbers of flares better distinguish a phased-flares signal from random flaring as described in Figure \ref{fig:snr_demo}. The ability to distinguish between periodogram power from phased flaring and power arising from random flaring is quantified by LS$_\mathrm{phased}$-LS$_\mathrm{random}$ and log $\mathcal{L}_\mathrm{phased}$-log $\mathcal{L}_\mathrm{random}$. These expressions give the typical increase in power from real detections versus non-detections. Intriguingly, the difference in power seen between phased and random flaring increases significantly in samples containing 3-24 flares, but the ability to discriminate a real signal from random flaring increases very modestly for samples larger than 24 flares. The difference in power from 3 to 74 flares is illustrated in the right panels of Figure \ref{fig:snr_demo}. Injected phased-flares are given a 3.142 d periodicity with a Gaussian 1$\sigma$ spread of 0.1 d while random flares are drawn from a uniform distribution of flare times sampled from the light curve.

\subsection{Identifying which stars may be phased-flare candidates}\label{ID_signal_methods}
For each of the 284 flare stars in our sample, the light curve and wide flares LS periodogram are plotted in separate panels of a graphical user interface (GUI). In a third panel, the light curve is folded in phase to the period at the highest peak of the flare periodogram. Stars with a clear peak in their periodogram or flares that appear to be regularly spaced in the light curve are recorded as potential candidates. Once the candidate periodicity is identified, the spread $\sigma_A$ in the preferred phase positions of each flare is measured and the threshold amplitude $A_\mathrm{thresh}$ is computed as described in \S \ref{ampl_thresh}. Potential candidates are not considered actual candidates until the false alarm probabilities are computed as described in \S \ref{fap_methods}.

\begin{figure*}
	\centering
	{
		\includegraphics[trim= 1 1 1 11,clip, width=6.9in]{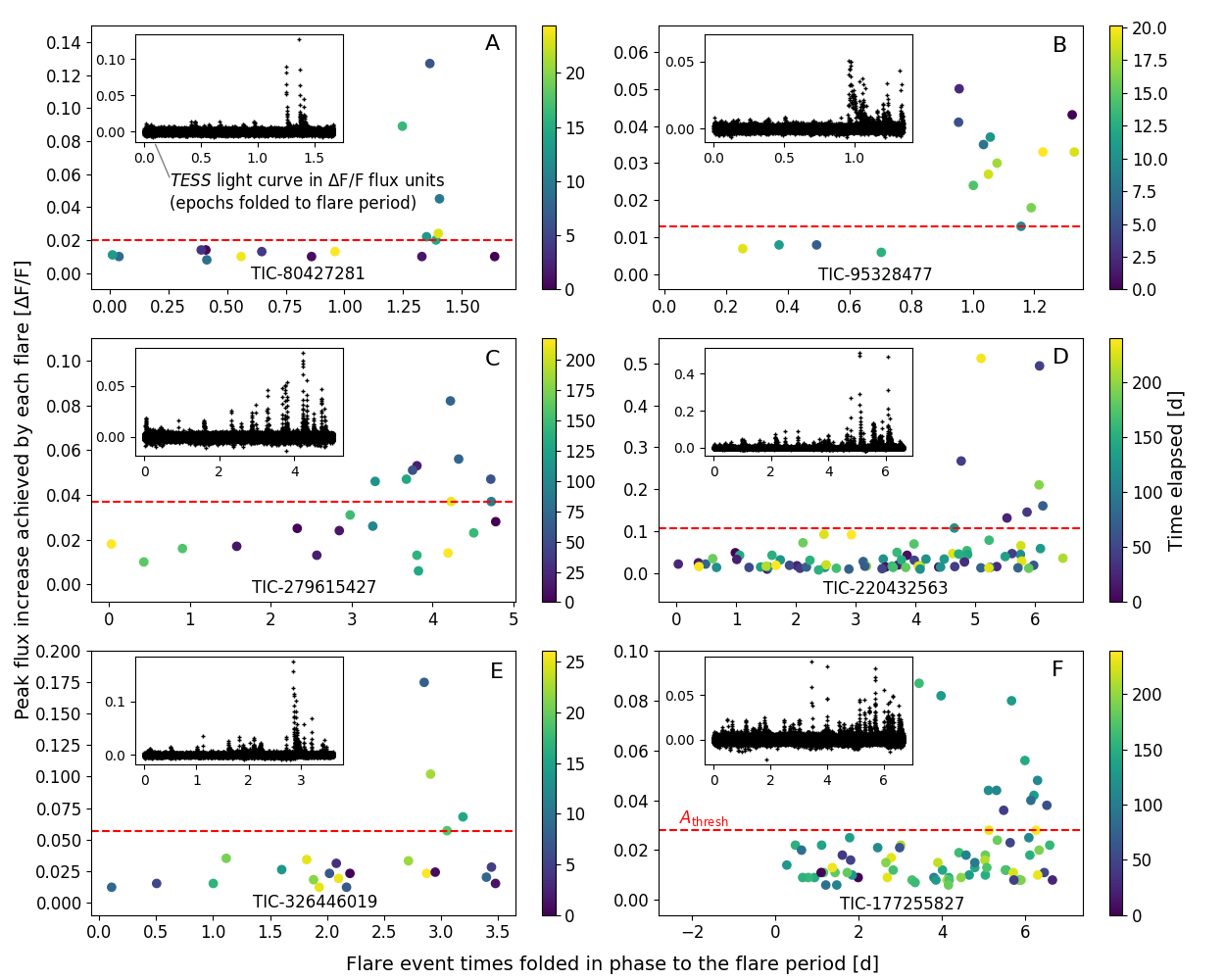}
	}
	\caption[Flare amplitudes versus phase-folded positions]{Larger panels: Flare flux amplitudes of each candidate versus flare times, folded in phase to the flare period. Flares are color-coded by position in the light curve to show whether groups of flares were emitted in the same day or different days. High amplitude flares are often emitted periodically at a particular phase. Smaller flares often appear to occur randomly. The dividing line is the threshold flare amplitude, $A_\mathrm{thresh}$. Inset panels: De-trended TESS light curves are folded in phase to the phased-flare period. The amplitude is in flux units, with the peak flux equivalent to the recorded amplitude values in the larger panels. For plotting purposes, the excess high-amplitude flares that occur at a particular periodic phase are plotted on the right-hand side of both larger and smaller panels.}
	\label{fig:ampl_phases_fig}
\end{figure*}
\subsection{Determining the threshold flare amplitude}\label{ampl_thresh}
For many of our targets, the largest flares in each light curve seem to exhibit the strongest periodicity. Unless the large flares can be separated from the smaller non-periodic flares, the period-power of the signal will be diluted. We therefore need a way to define a flare amplitude cutoff that distinguishes large flares from small non-phased flares. To avoid arbitrarily picking a threshold amplitude by hand, we define this threshold as the half-maximum of the $A$-$\sigma_A$ distribution (described below) and illustrated in the left panel of Figure \ref{fig:sigma_amplFWHM}. This is done as follows:

\begin{enumerate}
    \item The peak flare amplitudes $A$ are sorted from largest to smallest. The event times of each flare in the light curve are phase-folded to the candidate flare period.
    \item Starting with the largest flare, we select a sub-sample of flares $A_\mathrm{sub} = Ampl(A \geq A_\mathrm{cur})$ with amplitudes equal to or greater than that of the current flare amplitude $A_\mathrm{cur}$. The next step is to compute $\sigma_A$, the standard deviation of the event times of $A_\mathrm{sub}$ folded in phase to the candidate flare period as shown in the right panel of Figure \ref{fig:sigma_amplFWHM}.
    \item This process is repeated for each smaller flare in the sorted list, creating sorted ($A$, $\sigma_A$) pairs.
    \item The halfway point between the smallest and largest $\sigma_A$ values as located and shown in the left panel of Figure \ref{fig:sigma_amplFWHM}. The amplitude corresponding to the half-maximum of the $\sigma_A$ values is called $A_\mathrm{thresh}$.
    \item All flares are divided into two categories using the threshold (as shown in the right panel of Figure \ref{fig:sigma_amplFWHM}). Flares above $A_\mathrm{thresh}$ are selected as candidate phased-flare events and are subsequently used in periodogram detections and FAP calculations.
\end{enumerate}
Across all candidate phased-flare systems, we find $A_\mathrm{thresh}$ does a good job separating the strongly-phased flares from the lower-amplitude background flares.

\subsection{False-alarm Probabilities}\label{fap_methods}
We compute two different types of FAP on phased flares candidates. In the first type of FAP (denoted the phased-flare FAP), all flare times and amplitudes are randomly shuffled 10,000$\times$. In each trial, shuffled flares with amplitudes above $A_\mathrm{thresh}$ are selected and the standard deviation of the phased-up times $\sigma_A$ is computed. The FAP is defined as the fraction of trials where $\sigma_A$ is more tightly phased than the real signal. The benefit of this method is that it takes into account both the number of flares above $A_\mathrm{thresh}$ and also the degree to which the flares are tightly clustered in phase. For example, a low FAP may be due to a small number of extremely well-aligned flares in the phase-folded graph, or due to a larger number that all occur within half of the period but not in the other half. Both situations are observed in Section \ref{survey_results}. We cut all flare stars with a phased-flares FAP above 1\% from our sample as randomly-shuffled flares can too easily reproduce these signals. This cut flags the plausible candidates and keeps the number of false-positives to a manageable level in both the current and future larger survey of all TESS flare stars. In the current survey we would expect $\sim$3 false positives, and a full survey of all bright flare stars should result in 10-100 false positives (depending on the bright limit cutoff used).

The second type of FAP (denoted the periodogram FAP) describes how often random flaring reproduces detections in the two types of periodograms. For each detection, the number of flares with amplitudes above $A_\mathrm{thresh}$ is determined and an equivalent number of flare times are randomly drawn from the light curve. The Wide-flares and Bayesian-likelihood periodograms are computed across 10,000 MC trials for the randomly-drawn flares and the maximum power in the 1-10 d search range is recorded. The fraction of times the power at the period of interest is equal to or greater than the real signal gives the FAP. 

\begin{figure*}
	\centering
	{
		\includegraphics[trim= 1 25 1 10,clip, width=6.9in]{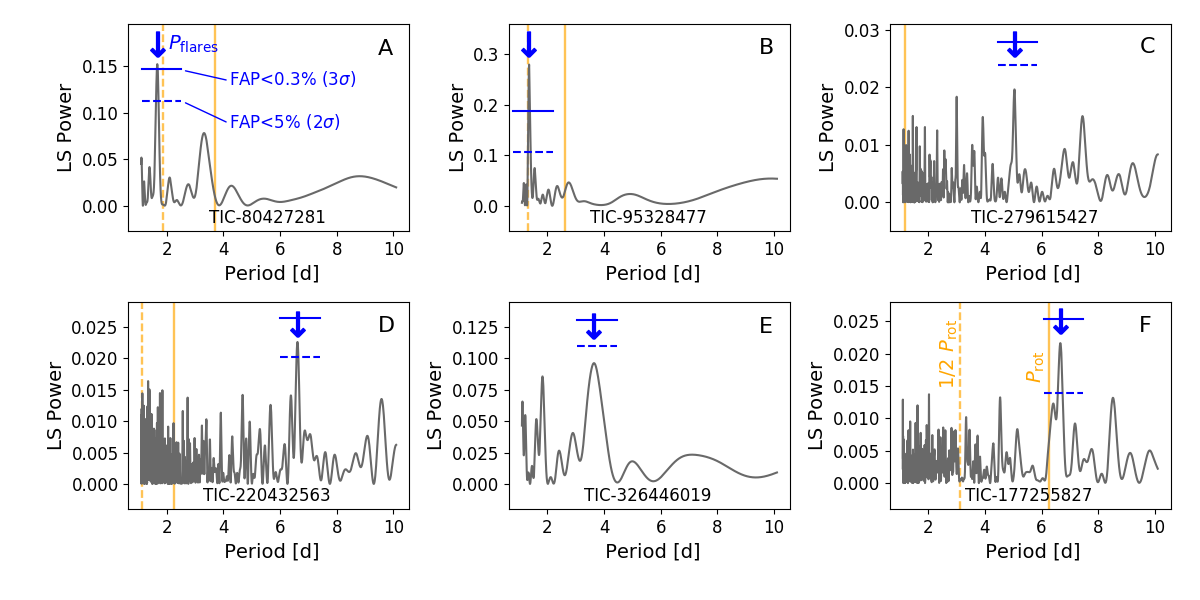}
	}
	\caption[Wide-flares Lomb-Scargle periodograms of each phased-flare candidate]{Wide-flares Lomb-Scargle periodograms of the strongest phased-flare candidates. Only candidates A,B, and D are detections. In each periodogram, the candidate $P_\mathrm{flares}$ is displayed with a blue arrow and is the most significant peak in its periodogram. Some peaks are clearer than others: for example the peaks in panels A and B are clearer than those in panels C and E. FAP are computed for each candidate period across 10,000 trials and are displayed in blue. For reference, the stellar rotation period is displayed as a solid orange line and the 1/2 alias as a dashed orange line.}
	\label{fig:WFLS_pgrams_fig}
\end{figure*}

We note this FAP assumes no prior exists on the number of periods searched. Sometimes the phased-flare period correlates with the stellar rotation period (e.g. peaks near a close harmonic of $P_\mathrm{rot}$ in Figures \ref{fig:WFLS_pgrams_fig} and \ref{fig:PFBL_pgrams_fig}), which provides a prior on the period search range. In such a case, the FAP is reduced by recording the maximum period per trial in a period range within the percent difference between the true phased-flares period and $P_\mathrm{rot}$ (or its closest harmonic). When the flare period is close to the $n$th harmonic of $P_\mathrm{rot}$, random power at $P_\mathrm{rot}$ and all harmonics up to $n$ is included in the FAP. For example, if the flare period is closest to 1/2$P_\mathrm{rot}$, then random periodogram power higher than the true phased-flares signal at $P_\mathrm{rot}$ and 1/2$P_\mathrm{rot}$ are both counted in each MC trial. The periodogram FAP will be higher than the phased-flares FAP since large LS peaks generated from 5-15 randomly-timed flares are easier to produce. The large range of the period search window will also increase these values relative to the single period tested in the phased-flares FAP.

Even given strong signals that pass our FAP checks, it is possible that autoregressive processes can lead to spurious periodicities. For example, stochastic auto-correlation and autoregressive behavior that could lead to pseudo-periodicity has been observed in the occurrence of solar flares, e.g. \citep{Lu_Hamilton1991, Crosby1993, Aschwanden2016, Stanislavsky2019}. While short-term spurious correlations can arise in principle, several additional factors work in favor of true periodicity in our sample. Each of our signals were discovered prior to measuring the stellar rotation period. While characterizing these signals (see \S \ref{survey_results}), we found the strongest signals in our data occur at the stellar rotation period or a harmonic of it. This is highly indicative of true periodicity rather than spurious correlation.

\section{Detection of phased flaring in TESS light curves}\label{survey_results}

We find 6 flare stars with candidate phased-flare signals in their TESS Cycle 1 data and a phased-flares FAP$<$1\%. These flare stars are TIC-80427281, TIC-95328477, TIC-279615427, TIC-220432563, TIC-326446019, and TIC-177255827. Only Candidates A (TIC-80427281), B (TIC-95328477), and D (TIC-220432563) are secure detections unlikely to be due to random peaks or look-elsewhere effects. Candidate F (TIC-177255827) is 2.5$\sigma$. The other two are non-detections (TIC-279615427 and  TIC-326446019). We note other statistical methods besides those we employed might report different significance values. Flare periods range from 1.3 to 6.7 d. 4 are M-dwarfs and 2 are likely late K dwarfs. We describe each detection below:
\begin{itemize}
  \item TIC-80427281 (ASAS J004211-4252.7): A moderately nearby (53 pc) M2.2 dwarf \citep{Kraus2014} observed for 1 sector by TESS in Cycle 1. We observe a flare period of 1.67 d with a phased-flares FAP of 0.1\%. This period is close to the half-alias of the stellar rotation period of $P_\mathrm{rot}$=3.715 d. Phase-folding the flares to $P_\mathrm{rot}$ does not reveal an obvious correlation (candidate A in the figures).
  \item TIC-95328477 (UCAC4 368-011078): A moderately nearby (51 pc) M0 dwarf \citep{Riaz2006} observed for 1 sector by TESS in Cycle 1. We observe a flare period of 1.35 d with a phased-flares FAP of 0.1\%. The stellar rotation period $P_\mathrm{rot}$=2.609 d is twice the phased-flare period. Phase-folding the flares to $P_\mathrm{rot}$ does show a clear sinusoidal modulation in flare amplitudes with rotational phase (candidate B).
  \item TIC-279615427 (UCAC3 63-21199): A moderately nearby (57 pc) K9 dwarf \citep{Howard_2020a} in the TESS Cycle 1 Continuous Viewing Zone (CVZ). We observe a flare period of 5.05 d with a phased-flares FAP of 0.4\%. The flare period does not correlate to the stellar rotation period $P_\mathrm{rot}$=1.1882 d and the flares do not show clear modulation when phase-folded to $P_\mathrm{rot}$ (candidate C).
  \item TIC-220432563 (2MASS J04534379-5836247): A moderately nearby (30 pc) K9 dwarf \citep{Howard_2020a} observed for 12 sectors by TESS in Cycle 1. We observe a flare period of 6.62 d with a phased-flares FAP of 0.1\%. The phased-flares period is $\sim$3X the $P_\mathrm{rot}$ of 2.274 d, making a correlation likely. However, phase-folding the flares to $P_\mathrm{rot}$ only shows a possible correlation between the largest flares and the dominant starspot (candidate D).
  \item TIC-326446019 (RBS 1877): A moderately nearby (35 pc) M3.5 dwarf \citep{Riedel2017} observed for 1 sector by TESS in Cycle 1. We observe a flare period of 3.63 d with a phased-flares FAP of 0.9\%. With a stellar rotation period $P_\mathrm{rot}$=0.8022 d, it is not likely the flare signal is correlated with the rotation period. Phase-folding the flares to $P_\mathrm{rot}$ shows no correlation either (candidate E). We also note that a plausible flare period near $P_\mathrm{rot}$ exists at 0.865 d, but this is below our constant minimum period search range and was found by tuning the periodogram search parameters. For consistency across all stars, we do not include this 0.865 d signal in our formal analysis but mention it here for completeness.
  \item TIC-177255827 (ASAS J064643-7700.4): A moderately nearby (64 pc) K7 dwarf \citep{Howard_2020a} in the TESS Cycle 1 CVZ. We observe a flare period of 6.68 d with a phased-flares FAP of 0.2\%. The stellar rotation period of $P_\mathrm{rot}$=6.26 d is nearly identical. While the flares phase up strongly at 6.68 d, they do not phase up well at 6.26 d, possibly due to differential rotation (candidate F).
\end{itemize}

\subsection{Phased-flaring at high amplitudes}\label{ampl_results}
We inspect the distributions of flare amplitudes versus their position in phase. Phase-folding the event times to the best flare period found with the periodograms, we observe that the flares cluster in phase as shown in Figure \ref{fig:ampl_phases_fig}. The threshold amplitude $A_\mathrm{thresh}$ separating large phased flares from small non-phased ones is shown in each panel. Some flare stars show a tighter clustering in phase than others. For example, the flares from Candidate A (TIC-80427281) occupy a very small range of positions in phase while the flares from Candidate F (TIC-177255827) cover almost half the phase. Large spreads in phase at periods close to the stellar rotation period as seen from Candidate F are likely indicative of stochastic flaring from active regions rotating into and out of the field of view. Alternately, they could be due to differential rotation and spot evolution. Very tight clustering may be indicative of triggering mechanisms \citet{Garcia_Alvarez2003, Gao2008}. These plots ensure $A_\mathrm{thresh}$ is accurately selecting the sample of flares for which the finalized periodograms are computed (i.e. those on which proper false-alarm calculations may be performed, not the initial search periodograms in high SNR flares).

In our sample, the high amplitude flares display the clearest signal. Flare periodicity amongst the highest-amplitude flares from a star has been detected in the past. \citet{Roettenbacher2013} found the brightest flares occurred when the dominant starspot was facing our line of sight, similar to our Candidate B (see panel B, Figure \ref{fig:p_rot_compare}). \citet{ Vida2016, Vida2017} found the strongest flares from V374 Peg and TRAPPIST-1 were concentrated at particular rotational phases, but not necessarily when the dominant spot was in our line of sight. 

Following \citet{Vida2016, Vida2017}, we also phase-fold the light curves themselves to the flare period. The phase-folded light curves are shown as inset panels in the phase-amplitude plots of Figure \ref{fig:ampl_phases_fig}. Each inset light curve is pre-whitened of non-flare variability with a SG filter and phase-folded to the flare period. The light curves help to ascertain at a glance if the amplitudes increase and decrease from a mean phase with the strongest flares as seen in Panel C, for example.

\subsection{Phased-flare periodogram results}\label{detection_pgrams}
\begin{figure*}
	\centering
	{
		\includegraphics[trim= 1 25 1 10,clip, width=6.9in]{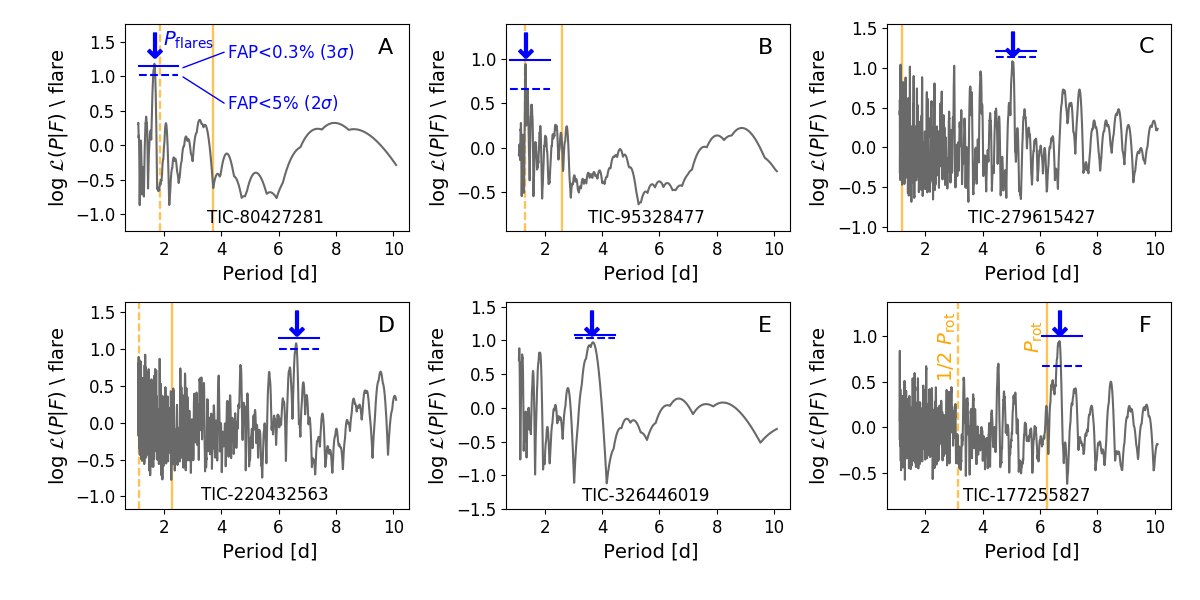}
	}
	\caption[Bayesian-likelihood periodograms of each phased-flare candidate]{Bayesian-likelihood periodograms of the 6 strongest phased-flare candidates. Only candidates A,B, and D are detections. In each periodogram, the signal $P_\mathrm{flares}$ is displayed with a blue arrow and is the most significant peak in its periodogram. Peaks are smaller than in the LS periodogram due to the logarithm used in computing periodogram power. Some peaks are clearer than others: for example the peaks in panels A and B are clearer than those in panels C and E. FAP are computed for each candidate period across 10,000 trials and are displayed in blue. For reference, the stellar rotation period is displayed as a solid orange line and the 1/2 alias as a dashed orange line.}
	\label{fig:PFBL_pgrams_fig}
\end{figure*}

Wide-flares LS and Bayesian-likelihood periodograms for each candidate are shown in Figure \ref{fig:WFLS_pgrams_fig} and Figure \ref{fig:PFBL_pgrams_fig}. In each periodogram, the phased-flare period is highlighted as the most significant detection in the period range searched. The detections are consistent between the LS and Bayesian likelihood approaches, although the peaks are higher in the LS periodograms. This is a result of the logarithm used in the Bayesian approach. The phased-flare periods are associated with either the stellar rotation period or an alias of it in 4/6 cases. This suggests the contrast of the flare against the stellar brightness changes as the emitting starspot rotates through our line of sight. 

The most secure detections are from Candidates A and B (TIC-80427281 and TIC-95328477), which are 3.3$\sigma$ and 4$\sigma$, respectively. The next strongest signals D and F (TIC-220432563 and TIC-177255827) are 3.3$\sigma$ and 2.5$\sigma$, respectively. Candidates C and E (TIC-279615427 and  TIC-326446019) are not detections. These $\sigma$ significance values are from the periodogram FAPs with constraints from stellar rotation included as described in Section \ref{fap_methods}. The FAP of candidate D is computed slightly differently from the others. The flare period is close to 3$P_\mathrm{rot}$, for a 2.3$\sigma$ signal (FAP=2.5\%). However, the rotational phases of the flares agree with the rotational phase of the dominant spot (at the light curve minimum) as shown in Figure \ref{fig:p_rot_compare}. We multiply the periodogram FAP by the probability that randomly-timed flares would correlate with the rotational phase of the dip. Across 10,000 MC trials, the root mean square error (RMSE) between the rotational phase of the dip and randomly timed flares is smaller than the RMSE of the real flares only 5.1\% of the time. This increases the signal strength to a 3.3$\sigma$ detection.

Across all 6 candidates, the least clear signals often have fewer flares with amplitudes above $A_\mathrm{thresh}$, although there are exceptions. For example, Candidate A has the second-strongest signal with only 6 periodic flares while the weaker Candidate C has 9 high-amplitude flares. The degree to which the phased flares are tightly-packed at a particular position in phase also affects the peak power. Candidate A's signal is relatively high because its flares are tightly clustered at a particular position in phase. To avoid unduly biasing our significance levels, we do not explore a large number of other statistical confirmation methods besides those described in this paper. Other methods besides those we employed might report different significance values.

\section{Periodicity related to starspot evolution}\label{stuff1}
We found each phased flares candidate in TESS Cycle 1 data, leaving Cycle 3 data for validation purposes and exploring the persistence of phased flaring through time. The Cycle 3 data is also leveraged to determine if phased flaring is caused by spot rotation, magnetic interaction with a companion, a flares reservoir, or the look-elsewhere effect.

\subsection{Spot evolution and possible periodicity in Cycle 3}\label{ruleout_spi}
\begin{figure*}
	\centering
	{
		\includegraphics[trim= 1 1 1 11,clip, width=6.9in]{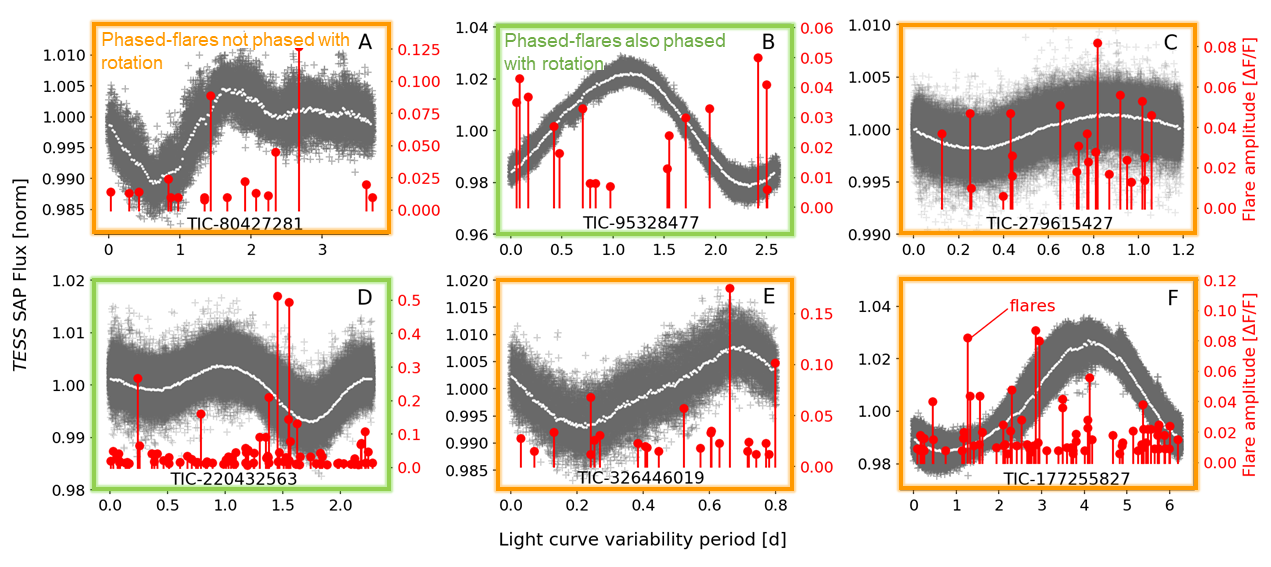}
	}
	\caption[Phased flares and TESS Cycle 1 Rotation Periods]{TESS Cycle 1 light curves are phase-folded to the stellar rotation period for each flare star and plotted in grey. Binned epochs are overlaid in white. Although it is difficult to distinguish between rotational modulation from multiple spots or EB scenarios in panels A and D, the phases and amplitudes of the variability are completely different a year later in Cycle 3, confirming the evolving starspot scenario. Flares are also phase-folded to the stellar rotation period and are shown in red. Only panels B and D show a convincing modulation in flare amplitude at the stellar rotation period. This does not mean other signals are not correlated with stellar rotation, just that they don't phase up well at the exact $P_{rot}$ value. Differential rotation at the spot latitude or external flare triggers could modulate phased-flares signals away from the precise $P_{rot}$ value.}
	\label{fig:p_rot_compare}
\end{figure*}

\begin{figure*}
	\centering
	{
		\includegraphics[trim= 1 1 1 11,clip, width=6.9in]{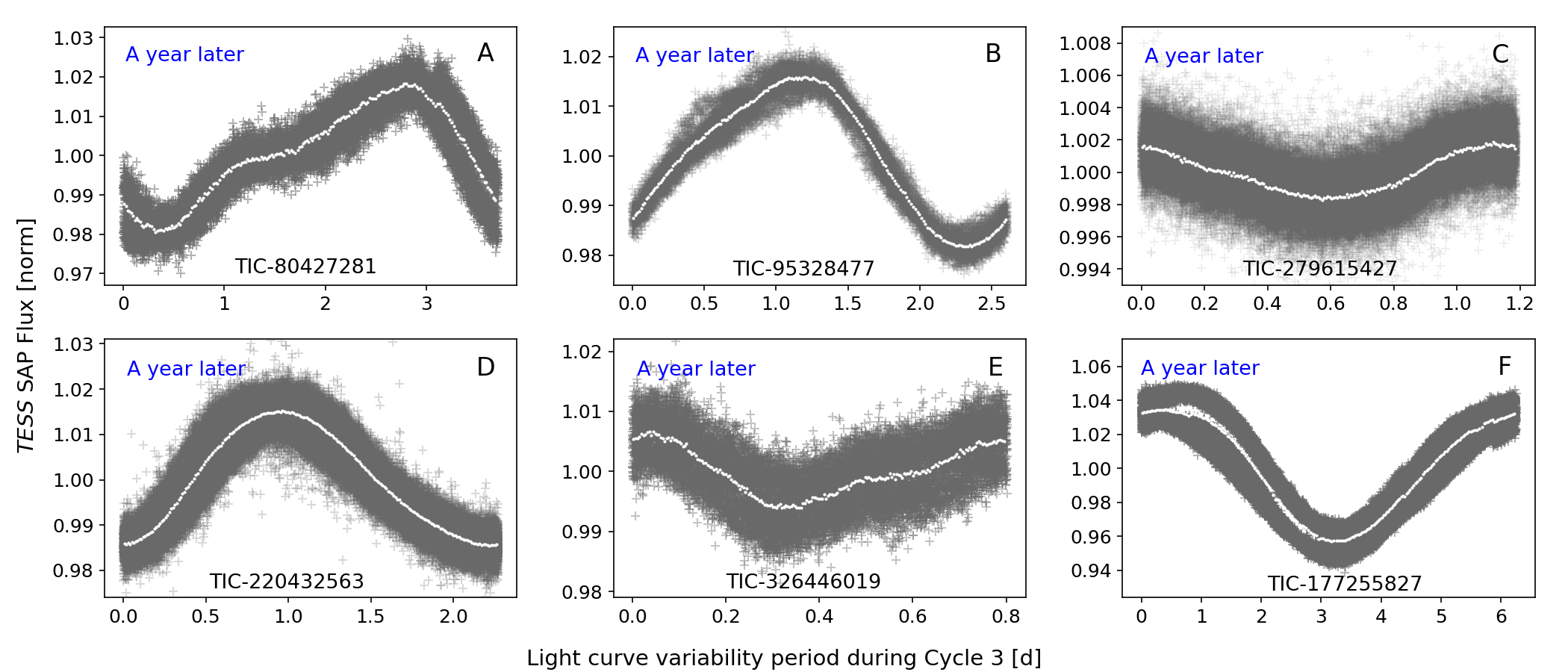}
	}
	\caption[Phased flares and TESS Cycle 3 Rotation Periods]{TESS Cycle 3 light curves are phase-folded to the stellar rotation period for each flare star and plotted in grey. Binned epochs are overlaid in white. Significant spot evolution has occurred in all stars but Candidate B since Cycle 1 a year earlier. This suggests magnetic properties generating flare periodicity may also have changed.}
	\label{fig:cycle3_p_rot}
\end{figure*}

None of the Cycle 3 data show strong signals at the exact flare periods found in Cycle 1 data, which is not surprising given spot evolution and differential rotation. Periodogram data give mixed results. The Cycle 3 data of Candidates B and F do show some support for periodicity near $P_\mathrm{rot}$ or 1/2 $P_\mathrm{rot}$. Candidate B also shows stronger support for a periodogram peak near twice the rotation period. The Cycle 3 data of A, C, D, and E do not show clear evidence of phased flaring. Candidate C appears less active in Cycle 3, with no flares above the Cycle 1 $A_\mathrm{thresh}$ amplitude cutoff. Several of the other targets also appear less active, but this could be an effect of Poisson flare statistics. The variability in the periodogram results and lack of periodicity entirely at the exact periods found in Cycle 1 could indicate the signals depend on magnetic properties that may have changed through time, or this could indicate some signals were a result of the look-elsewhere effect. Intriguingly, the rotational modulation seen in the Cycle 3 light curves of Figure \ref{fig:cycle3_p_rot} has evolved significantly from the modulation seen in the Cycle 1 light curves of Figure \ref{fig:p_rot_compare}. The difference is especially pronounced for the multi-spot rotators in panels A and D in Figure \ref{fig:p_rot_compare}. Candidate D shows 2 dips in Cycle 1 reminiscent of a W UMa, but in Cycle 3 shows only a single dip. The changes in number of dips and the amplitudes and phases of variability strongly indicates each star is undergoing significant spot evolution as in \citet{Roettenbacher2018}. The similarity of our samples may explain why we also find flare periodicity.

\subsection{Could the look-elsewhere effect be the cause of periodicity?}\label{look_elsewhere}
Could the potential lack of clear periodicity in Cycle 3 mean the Cycle 1 signals are due to the look-elsewhere effect? The look-elsewhere effect (also known as the multiple testing problem) is not a plausible explanation for the secure detections (candidates A, B, and D) but is plausible for the unconfirmed detections. For example, the probability that our strongest detection (Candidate A) could arise from the look-elsewhere effect in a survey of 300 stars is 2.3\% (based on the phased-flares FAP). The probability that both Candidates A and B arise from the look-elsewhere effect is $\sim$0.1\%. 

In each of 10,000 MC trials, we search 300 flare stars for stronger phased flaring than the target signal. The strength of phased flaring in each trial is defined as the degree of clustering at a particular position in periodic phase as described in \S \ref{fap_methods}. The number of stars in the sample with a tighter clustering than that of the actual flare star(s) is recorded as a ``success", and the fraction of trials with a success is given as the final look-elsewhere probability. 

However, the look-elsewhere effect cannot be excluded as an explanation of periodicity for candidates C, E, or F. Candidate F is the most likely to be astrophysical since it correlates with the stellar rotation period, but this probability is difficult to quantify: If the phased-flares FAP is used, the chance of a signal like F resulting from the look-elsewhere effect in a sample of 300 stars is $\sim$5\%. However, if the periodogram FAP is used, a 2.5$\sigma$ event should occur by chance $\sim$4 times in a sample of 300 stars. Individual 2.5$\sigma$ candidates near the rotation period or another period of interest (such as a candidate at a period expected from SPI) may still be astrophysical and should be investigated.

We note that our approach to the look-elsewhere effect may be overly conservative. We employ family-wise error rate (FWER) correction, which minimizes false positives at the expense of missing weak but real signals. Methods such as the False Discovery Rate (FDR; \citealt{Benjamini1995}) accept some false positives in order to provide greater statistical power than our FWER approach in detecting weak signals. It is possible that FDR could detect the marginal signals in our sample missed by FWER.

\subsection{Companion stars and evolving spot properties}\label{spot_explanations}
Since the look-elsewhere effect is not the best explanation for the periodicity of our best candidates in Cycle 1, we consider astrophysical interpretations. Since the periodicity changed when the spot structures present in the light curves changed, a connection between the two is probable. The phased-flare signals from candidates A, B, D, and F indicate rotational modulation because they are close to either $P_\mathrm{rot}$, the 1/2 $P_\mathrm{rot}$ harmonic, or the 3 $P_\mathrm{rot}$ harmonic. A 3 $P_\mathrm{rot}$ harmonic might result from delayed flare events or a juxtaposition of the effects of the wait-time between flares and $P_\mathrm{rot}$. Two show more power at the 1/2 alias than the primary period, likely a result of the basis functions of the periodograms. Candidate B in particular shows a clear modulation of amplitudes with rotational phase (Figure \ref{fig:p_rot_compare}). When the dominant spot is facing us at light curve minimum, we observe the largest flares. However the smallest flares occur at light curve maximum. Likewise, the largest flares correlate with the dominant spot in panel D of Figure \ref{fig:p_rot_compare}.

As seen in Figure \ref{fig:candidate_lightcurves_fig}, the periodicity of our clearest candidates prefers a very small range of rotational phases. As the spot rotates into and out of our line of sight, the largest flares should be observed at phase positions covering up to half the period. Furthermore, the preference of candidates A and B for two phase positions separated by half a stellar rotation period is suggestive of the phased flaring detected from the binaries YY Gem and V711 Tau by \citet{Doyle1990} and \citet{Garcia_Alvarez2003}. RVs would be necessary to rule out a similar situation here since binarity is not evident in the TESS light curves. Spot evolution is observed in each light curve via changes in the amplitudes and phases of rotational variability (with the possible exception of Candidate B). Spot evolution is especially pronounced in Candidate F, making reflection/ellipsoidal modulation highly unlikely. The lack of phased flaring in Cycle 3 could also suggest there is no external trigger such as a companion star. However, if both a companion star and the right magnetic conditions are both required for periodicity to occur, then phased flaring could come and go as spots evolve. If no companion star or planet is forcing the periodic release of magnetic energy, it is possible that conditions in the dominant spot may create a temporary flare reservoir. We note this possibility is highly speculative.

\section{Conclusions}\label{disc_conclude}
We have developed and tested two new approaches to detect and statistically confirm flare periodicity. The periodograms are leveraged to perform the first large-scale survey for periodicity in flare rates that explores periods other than $P_{rot}$. We confirm 3 detections, which should also be targeted with RV to look for non-transiting companions and planets. After imposing the initial 1\% FAP cut (not the 0.3\% cut required for final detection), at least 2\% of flare stars are found to exhibit possible periodicity. This suggests at least $\sim$20 detections should occur in a survey of the brightest 2 min cadence TESS light curves of late K and M-dwarf stars (assuming the fractions of flare stars from \citet{Howard_2019}). 

Should flare periodicity be shown to probe SPIs as predicted by \citet{Lanza2018}, TESS light curves will open a new window on the discovery and characterization of close-in exoplanets. For example, the magnetic fields of M-dwarf terrestrial planets remain unexplored \citep{Luger2017}. Probing their magnetospheres through phased flares might potentially enable the first tests of magnetic dynamo evolution of rocky planets outside our solar system and inform planetary habitability. Alternately, a search for flare periodicity across the entire TESS data set would place two orders of magnitude better constraints on the occurrence of flare reservoirs or interacting binaries like YY Gem than has previously been possible. Building on the work of \citet{Feinstein2020}, the conditions under which TESS flares depend on active longitude could be ascertained.

Our survey has demonstrated the feasibility of extending our flare periodogram analysis to the brightest 15,000 cool stars that have high cadence TESS light curves. The amplitudes of flares sufficiently large to be part of a candidate phased-flares signal are robustly determined by the $A$-$\sigma_A$ mechanism. Once the sample of candidate flares has been identified, Monte-Carlo calculations of periodogram power and signal strength are obtained and the causes of the periodicity may be explored.

Such a survey will often detect correlations between the phased flares period and the stellar rotation period. We suspect that SPI signals may be difficult to identify and/or rare. We note it is important to consider quasi-periodicities arising from autoregressive and stochastic auto-correlative behavior in assessing the causes of apparent periodicity (i.e. such as SPI). Furthermore, flare periodicity may be transitory. This implies separate years of data from the same star may need to be analyzed independently. Our results strengthen the correlation between flare periodicity and spot evolution from \citet{Roettenbacher2018}.

\section*{Acknowledgements}\label{acknowledge}
WH thanks Andrew Mann and Chris Clemens for helpful office conversations on flare periodicity and Adina Feinstein for a helpful Twitter conversation on flare periodicity.
\par WH acknowledges funding support by the TESS Guest Investigator Program GO 3174.
\par This paper includes data collected by the TESS mission. Funding for the TESS mission is provided by the NASA Explorer Program.
\par WH and NL acknowledge the band Gloryhammer for extra motivation to write.
\par This work has made use of data from the European Space Agency (ESA) mission {\it Gaia} (\url{https://www.cosmos.esa.int/gaia}), processed by the {\it Gaia} Data Processing and Analysis Consortium (DPAC, \url{https://www.cosmos.esa.int/web/gaia/dpac/consortium}). Funding for the DPAC has been provided by national institutions, in particular the institutions participating in the {\it Gaia} Multilateral Agreement.
\par This research made use of Astropy,\footnote{http://www.astropy.org} a community-developed core Python package for Astronomy \citep{astropy:2013, astropy:2018}, and the NumPy, SciPy, and Matplotlib Python modules \citep{numpyscipy, 2020SciPy-NMeth, matplotlib}.

{\it Facilities:} \facility{NASA:TESS}

\bibliographystyle{apj}
\bibliography{paper_references}

\end{document}